\documentclass[12pt]{article}
\usepackage{epsfig,graphics,graphicx}
\usepackage{amssymb}
\usepackage{amsmath}
\usepackage{extarrows}
\usepackage{algorithm}
\usepackage{algpseudocode}
\usepackage[utf8]{inputenc}
\usepackage[english]{babel}
\usepackage{geometry}
\usepackage{amsthm}
\usepackage{scalerel}[2014/03/10]
\usepackage[usestackEOL]{stackengine}
\usepackage{esint}
\usepackage{xcolor}
\usepackage{authblk}
\def\Z{\mathbb Z}

\title{\sc{Introduction to Number Theoretic Transform}}
\date{}

\title{\sc{Revisit to the Bai-Galbraith signature scheme}}

\author[1, 2]{Banhirup Sengupta}
\author[1]{Peenal Gupta}
\author[3]{Souvik Sengupta}

\affil[1]{Research Group, PinakashieldTech OÜ, Tallinn, Estonia}
\affil[2]{Centre for Applicable Mathematics, Tata Institute of Fundamental Research, Bangalore, India}
\affil[3]{Digital Ecosystems, IONOS SE, Karlsruhe, Germany}

\begin{document}
	
	\maketitle
	
	\abstract{Dilithium is one of the NIST approved lattice-based signature schemes. In this short note we describe the Bai-Galbraith signature scheme proposed in \cite{BG}, which differs to Dilithium, due to the fact that there is no public key compression. This lattice-based signature scheme is based on Learning with Errors (LWE).}
	
	\section{Introduction}
	Lattice-based cryptography is considered as a promising alternative to traditional cryptography after the introduction of quantum computers. An important problem is to obtain practical/provably secure public key signature schemes based on lattice assumptions. One approach is to use trapdoor functions and the hash-and-sign methodology (see \cite{GPV} and \cite{SS}). However, the Fiat-Shamir paradigm has been the most promising path for
	practical signatures. A series of works by Lyubashevsky and others \cite{DDLL}, \cite{GLP}, \cite{L}, \cite{LY}, have developed schemes based on the Fiat-Shamir paradigm that are secure in the random oracle model. When implementing lattice-based signature
	schemes, one might face several challenges. For instance, the size of the public key and that of the signature. Furthermore, the additional
	requirement to sample from discrete Gaussians during the signing process. However, the Bai-Galbraith scheme mainly focuses on the reduction of the size of signatures. \\
	The core idea of Lyubashevsky's signatures in case of LWE is to have a public key of the form $(A,T= AS + E (\mod q))$ where $A$ is an $m\times n$ matrix and $m\approx n$. The signing process commences by choosing vectors $y_{1}$, $y_{2}$ of small norm and computing $v = A y_{1} + y_{2} (\mod q)$. Next, the signer computes $c = H(v, \mu)$, using the Fiat-Shamir
	paradigm, where $\mu$ is the message and $H$ is a
	hash function. Finally, the signer computes $z_{1} = y_{1} +Sc$ and $z_{2} = y_{2} +Ec$. The signature is $(z_{1}, z_{2}, c)$. The verifier checks that $\|z_{1}\|$ and $\|z_{2}\|$ are small enough and that $H(Az_{1} + z_{2} - Tc ( \mod q), \mu)$ is equal to $c$. A significant obstacle to
	short signatures is the need to send the length $m$ vector $z_{2}$. The main novelty of the Bai-Galbraith signature scheme is that $z_{2}$ can be omitted entirely. Lyubashevsky's scheme \cite{GLP}, \cite{LY},
	based on LWE has public key $(A, b = As+e ( \mod q))$ and a signature is like a proof of knowledge of the pair $(s, e)$. The key feature of B-G scheme is to prove the knowledge of $s$ alone. The smallness of $e$ becomes implicit in the verification equation. So, there is no need to send any information about $e$. Since $s$ has length $n$
	and $e$ has length $m\approx n$, not requiring to prove knowledge of $e$ has the potential to provide a significant reduction in signature size. \\
	One of the drawbacks of B-G scheme is the use of standard LWE instead of
	Ring-LWE. Lattice-based signatures schemes using Ring-LWE or NTRU would give more practical signatures. While there are certainly significant practical benefits from using Ring-LWE (such as smaller public keys), there are also some constraints (such as considering $n$ to be an exponent of $2$), and so the Ring-LWE case is a little less flexible. To conclude, the two advantages of using standard LWE are: we get security based on standard assumptions in general lattices; we have complete flexibility in the parameters $(n,m)$ for LWE
	(instead of being stuck with $m = 2n$ and $n = 2^{d}$). \\
	This note is mostly taken from \cite{BG} and \cite{DET}. 
	
	\section{Algorithmic representation}
	Algorithm~\ref{alg:personalized_dil} presents the modified version of the Bai-Galbraith scheme introduced in \cite{BG}. \\
	The Key generation algorithm generates a  $k\times l$ matrix $A$. Each of the entries of $A$ is a polynomial in the cyclotomic ring $R_q = \Z_q[X]/(X^{n} +1)$. We consider $q = 2^{23}-2^{13} + 1 $ and $n = 256$. The algorithm samples random secret key vectors $s_1$ and $s_2$. Each coefficient of these vectors is an element of $R_q$ with
	small coefficients, having size at most $\eta$. Finally, the second part of the public key is computed as $t = A s_1 + s_2$. The algebraic operations in this scheme are performed over the polynomial ring $R_q$. \\
	The signing algorithm generates a vector of polynomials $y$ with coefficients equal to $\gamma$. The parameter $\gamma$ is set in a strategic manner. It is large enough so that the eventual signature does not reveal the secret key, yet small enough so that the signature is not easily forged. The signer then computes $Ay$ and sets this as $w$. Next, the challenge polynomial $c$ is created as the hash of the message and the high-order bits of $w$. The reason behind this distribution is the fact that $c$ has small norm and comes from a domain of size greater than $2^{256}$. The potential signature is then computed as $z = y + c s_1$. \\
	At this point, if $z$ were directly output, then the signature scheme would be insecure	due to the fact that the secret key would be leaked. We use rejection sampling in order to avoid the dependency of $z$ on the secret key. The parameter $\beta$ is set to be the maximum possible coefficient of $c s_{i}$. Since $c$ has $60$ $+1$'s and $-1$’s and the maximum coefficient in $s_{i}$ is $\eta$, it is easy to check that $\beta\leq 60\eta$. If any coefficient of $z$ is larger than $\gamma-\beta$, then we reject and restart the signing procedure. Furthermore, if any coefficient of the low-order bits of $A z - ct$ is greater than $\gamma-\beta$, we restart. The first check is necessary for security, while the second is necessary for both security and correctness. \\
	In the verification part, the verifier computes the high-order bits of $A z - ct$, and then accepts if all the coefficients of $z$ are less than $\gamma-\beta$ and $c$ is the hash of the message and the high-order bits of $A z - ct$. To make sure that the verification works, we need to check that $High(Az-ct)=High(Ay)$. One must note that $Az-ct=Ay-cs_{2}$. So all we need to show is that
	\begin{equation}\label{high}
		High(Ay) = High(Ay-cs_{2}).
	\end{equation} 
	The reason for this is that a valid signature will have $\|Low(Ay-cs_{2})\|_{\infty}<\gamma-\beta$. Since we know that the coefficients of $cs_{2}$ are smaller than $\beta$, we know that adding $cs_{2}$ is not enough to cause any carries by increasing any low-order coefficient to have magnitude at least $\gamma$. Hence, Equation \eqref{high} is true and the signature is correctly verified.
	
		\begin{algorithm}
		\caption{Modified Bai-Galbraith Scheme}
		\label{alg:personalized_dil}
		\resizebox{\textwidth}{0.3\textheight}{ 
			\begin{minipage}{\textwidth}
				\begin{algorithmic}[1]
					
					\State \textbf{Key generation:}
					
					\State $A \leftarrow R_{q}^{k \times l}$
					\State $(s_{1}, s_{2}) \leftarrow S_{\eta}^{l} \times S_{\eta}^{k}$
					\State $t := A s_{1} + s_{2}$
					\State return $(pk = (A,t), sk = (A,t, s_{1}, s_{2}))$
					
					\State \textbf{Signing:}
					\State $y \leftarrow S_{\gamma}^{l}$
					\State $w=Ay$
					\State $c = H(High(w),M)\in B_{60}$
					\State $z = y + c s_{1}$
					\If{$\|z\|_{\infty} > \gamma - \beta$ or $\|Low(w-cs_{2})\|_{\infty}>\gamma - \beta$}
					\State restart 
					\Else
					\State return $sig =(z,c)$
					\EndIf
					\State \textbf{Verification:}
					\State $c'= H(High(Az-ct),M)$
					\If{$\|z\|_{\infty}\leq \gamma - \beta$ and $c'=c$}
					\State return accept
					\Else
					\State return reject
					\EndIf
				\end{algorithmic}
			\end{minipage}
		}
	\end{algorithm}

\section*{Conclusion}

The Bai-Galbraith (B-G) signature scheme presents a compelling approach to post-quantum cryptography, specifically in the context of lattice-based signatures. A primary challenge in implementing these schemes is managing the size of the public key and the signature itself. The B-G scheme addresses this by omitting the need to send a vector typically included in the signature, thereby providing a significant reduction in signature size. This reduction has direct implications for data security and privacy, as smaller signatures can be transmitted more efficiently, reducing bandwidth and storage requirements, which is particularly beneficial for resource-constrained environments like embedded systems.

The security of the B-G scheme is rooted in the standard Learning with Errors (LWE) problem, a well-studied assumption in general lattices. Unlike schemes that rely on Ring-LWE or NTRU, the use of standard LWE offers flexibility in parameter selection, such as the matrix dimensions $(n,m)$, without being constrained to specific values like $n=2^d$. This flexibility allows for a broader range of security-performance trade-offs, enabling developers to tune the scheme for specific privacy and security needs. The core innovation of the B-G scheme lies in proving knowledge of the secret key 's' alone, with the smallness of the error vector 'e' being implicitly verified. This clever technique avoids the need to transmit information about 'e', which directly contributes to the smaller signature size and, by extension, enhances data privacy by minimizing the data revealed during the signing process.

The signing process incorporates a rejection sampling technique, which is crucial for preventing the leakage of the secret key and maintaining security. This mechanism ensures that the final signature does not depend on the secret key in a way that could compromise it. This is a critical security feature, as it protects the integrity of the private key, which is the foundation of the signer's identity and data authenticity. Ultimately, the Bai-Galbraith scheme demonstrates that it's possible to design an efficient and provably secure lattice-based signature scheme that addresses key practical challenges, offering a promising path forward for securing data in the post-quantum era.

\end{document}